\documentclass[10pt]{article}
\usepackage{iclr2025_conference}
\usepackage{amsmath,amssymb,mathtools}
\usepackage{booktabs}
\usepackage{multirow}
\usepackage[ruled,vlined,linesnumbered]{algorithm2e}
\usepackage{hyperref}
\usepackage{xcolor}
\usepackage{microtype}
\usepackage{enumitem}
\usepackage{graphicx}
\usepackage[numbers,sort&compress]{natbib}
\usepackage{subcaption}
\usepackage{float}

\hypersetup{colorlinks=true,linkcolor=blue!60!black,citecolor=blue!60!black,urlcolor=blue!60!black}

\newcommand{\R}{\mathbb{R}}
\newcommand{\bx}{\mathbf{x}}
\newcommand{\bh}{\mathbf{h}}
\newcommand{\bz}{\mathbf{z}}
\newcommand{\br}{\mathbf{r}}
\newcommand{\bW}{\mathbf{W}}
\newcommand{\bq}{\mathbf{q}}
\newcommand{\bQ}{\mathbf{Q}}
\newcommand{\bK}{\mathbf{K}}
\newcommand{\bV}{\mathbf{V}}
\newcommand{\bM}{\mathbf{M}}
\newcommand{\bA}{\mathbf{A}}
\newcommand{\bp}{\mathbf{p}}
\newcommand{\bC}{\mathbf{C}}
\newcommand{\cL}{\mathcal{L}}

\title{ASR-Agnostic Multimodal Spectrotemporal Modeling\\for Early Dementia Detection}

\author{Ugwu Chukwuemeka and Richard Oluwafemi Oyeleke}

\date{}

\begin{document}
\maketitle

\begin{abstract}
Speech production recruits the same executive, attentional, and working memory processes that underpin instrumental activities of daily living (IADLs), making it a compelling non-invasive proxy for functional cognitive assessment. Yet most speech-based dementia detection systems depend on transcription, discard within-recording temporal structure, and are validated on a single English corpus with documented recording artifacts. We propose an ASR-agnostic framework that operates directly on Mel spectrograms. Our key contribution is the extraction of \textbf{spectrotemporal displacement fields} from consecutive spectrogram frames, capturing how spectral energy patterns shift and fluctuate as digital biomarkers of cognitive decline. These features are fused with CNN--ConvGRU acoustic embeddings through a \textbf{learned cross-attention mechanism} and aggregated via a Transformer encoder with learnable query pooling. A composite temporal loss enforces smoothness and contrastive coherence across segments. We train three independent models on English (DementiaBank), Slovak (EWA-DB), and Spanish (Ivanova) corpora, each grounded in clinical cognitive elicitation protocols that tax IADL-relevant cognitive domains. The Slovak model achieves 83.9\% accuracy (F1\,=\,0.878), Spanish achieves AUC\,=\,0.788, while the English baseline yields 53.2\%, confirming known corpus artifacts. Ablation studies across all three languages reveal three distinct fusion regimes: removing cross-attention collapses Spanish performance to 53.7\% (below either unimodal model), while on Slovak the audio encoder alone outperforms the full model (93.7\% vs.\ 83.9\%), and all English configurations remain near chance. This demonstrates that the value of multimodal fusion is corpus-dependent: essential when signal is distributed across modalities, counterproductive when one modality dominates, and irrelevant when no signal exists. Auxiliary temporal losses converge to language-invariant values (feature CV\,=\,3.4\%), providing evidence of cross-lingual architectural stability.
\end{abstract}

\section{Introduction}
\label{sec:intro}

Dementia affects over 55 million people worldwide, with cases projected to reach 153 million by 2050~\citep{nichols2022gbd}. Diagnosis currently depends on amyloid PET imaging, cerebrospinal fluid analysis, and multi-hour neuropsychological batteries, all of which are expensive, invasive, or inaccessible. Up to 75\% of dementia cases remain undiagnosed globally, rising to 90\% in low- and middle-income countries~\citep{who2021dementia,livingston2024lancet}. This diagnostic gap has motivated interest in speech as a non-invasive screening modality~\citep{kourtis2019digital,qi2025landscape}. The clinical logic is compelling: speech production is not merely a language task but a complex cognitive act that simultaneously recruits executive function, working memory, attention, and self-monitoring, the same domains whose decline defines impairment in instrumental activities of daily living~\citep{marshall2011executive,razani2007executive}. Verbal fluency, for instance, correlates more strongly with observation-based functional tests than any other cognitive measure in mild dementia~\citep{razani2007executive}, and connected speech decline has been shown to precede clinical MCI diagnosis~\citep{mueller2018connected}. The elicitation tasks used in our three evaluation corpora (picture description, structured naming, reading aloud) are specifically designed to tax these overlapping cognitive demands, positioning speech analysis as a continuous, ecologically valid proxy for IADL-relevant function~\citep{lanzi2023dementiabank,forbes2005detecting}.

Three limitations constrain the current landscape. First, the majority of systems depend on automatic speech recognition (ASR) or manual transcription to extract linguistic features~\citep{shakeri2025nlp}, introducing a dependency on transcription quality that degrades across languages and recording conditions. Purely acoustic approaches avoid this but plateau at 70--82\% accuracy~\citep{haider2020paralinguistic}. Multimodal fusion has pushed results to 83--93\%~\citep{ilias2022multimodal,lin2024scientific}, yet naive fusion provably \textit{hurts}: modalities compete under joint training, and the weaker modality may never be discovered~\citep{wang2020cvpr,huang2022modality}. Second, most systems aggregate features across an entire recording, discarding within-utterance temporal evolution. Dynamic modeling outperforms static aggregation by roughly 8 percentage points~\citep{syed2020static}, and clinical evidence confirms that dementia manifests as temporal instabilities in jitter, shimmer, pitch contour, and speech rhythm~\citep{parlak2023voice,meilan2020rhythm}. Third, 63\% of speech-based AD studies rely on DementiaBank~\citep{shakeri2025nlp}, a corpus in which classification models achieve near-perfect accuracy using only silent segments, exploiting recording artifacts rather than speech biomarkers~\citep{liu2024cleverhans}.

We address all three limitations through a unified framework with four contributions:

\begin{enumerate}[leftmargin=*,itemsep=2pt,topsep=4pt]
    \item \textbf{Spectrotemporal displacement fields.} We compute dense 2D displacement fields between consecutive spectrogram frames, capturing formant trajectory deviations, pitch instability, and energy redistribution as digital biomarkers. This extends the spectrogram-as-image paradigm~\citep{gong2021ast} and 1D audio flow~\citep{ezzat2005audioflow} to full 2D spectrotemporal analysis.
    \item \textbf{Cross-attention multimodal fusion.} Spectral dynamics features are fused with CNN--ConvGRU acoustic embeddings through learned cross-attention, preventing the destructive modality interference documented under naive combination~\citep{huang2022modality}.
    \item \textbf{Temporal regularization.} A composite loss enforces smoothness, contrastive coherence, progression regularity, multi-scale consistency, and attention sparsity across speech segments.
    \item \textbf{Trilingual validation with complete ablation.} Three independent models trained on English, Slovak, and Spanish corpora whose elicitation protocols tax IADL-relevant cognitive domains. Full ablation on all three languages reveals three distinct fusion regimes, establishing that the value of cross-attention is corpus-dependent rather than universally beneficial.
\end{enumerate}

\section{Related Work}
\label{sec:related}

\subsection{Speech as a Cognitive and Functional Biomarker}

The theoretical basis for speech-based dementia detection rests on the shared cognitive architecture between speech production and daily functional competence. Levelt's model decomposes speech into conceptualization, formulation, and self-monitoring~\citep{levelt1989speaking}, each requiring the executive, attentional, and memory resources that IADLs also demand~\citep{kempler2010language}. Marshall et al.\ showed that executive dysfunction independently predicts IADL decline even after controlling for memory~\citep{marshall2011executive}, and connected speech has been validated as a marker of disease progression in autopsy-confirmed AD~\citep{ahmed2013connected}. De la Fuente Garcia et al.\ reviewed 51 studies and identified the Pitt Corpus as the field's most heavily used resource, while arguing for ecologically valid monitoring paradigms~\citep{delafuente2020systematic}. Recent work has begun validating remote, home-based speech collection with associations to amyloid pathology~\citep{vandenberg2024remote} and longitudinal cognitive tracking~\citep{side2024longitudinal}.

\subsection{ASR-Agnostic and Acoustic Approaches}

Agbavor and Liang achieved AUC of 0.846 using data2vec without any transcription~\citep{agbavor2023endtoend}. Haider et al.\ showed that paralinguistic features alone reach 78.7\% accuracy through hard fusion~\citep{haider2020paralinguistic}. The ADReSS-M Challenge confirmed that speech timing features (pause duration, speech rate) transfer across languages with AUC\,=\,0.75, while lexical features fail~\citep{luz2024adressm,pereztoro2025crosslingual}. These results motivate our fully ASR-agnostic design, operating on Mel spectrograms without transcription or language models.

\subsection{Multimodal Fusion and Its Failure Modes}

Multimodal dementia detection has shown promise: Chu et al.\ achieved AUC of 93.0\% with audio-visual fusion~\citep{chu2023audiovisual}, and Lee et al.\ used Shapley-value weighting across three modalities for 90.6\% accuracy~\citep{lee2025multimodal}. However, Wang et al.\ demonstrated that naive late-fusion networks underperform unimodal baselines~\citep{wang2020cvpr}, Huang et al.\ proved this modality competition theoretically~\citep{huang2022modality}, and Du et al.\ showed that jointly trained encoders learn worse representations than independent ones~\citep{du2023unimodal}. Our cross-attention mechanism addresses this by enabling selective, temporally-aligned information exchange rather than blind combination. Importantly, all prior multimodal systems require separate sensor streams; our framework derives both modalities from the same audio signal.

\subsection{Temporal Modeling and Spectrotemporal Analysis}

Pan et al.\ used path signatures to preserve sequential speech patterns~\citep{pan2023path}. Gao et al.\ proposed dual-stage temporal attention~\citep{gao2025dstcnet}. Our predecessor method adapted RAFT's iterative refinement for spectrograms but explicitly did not compute displacement vectors~\citep{ugwu2025taispeech}. Ezzat et al.\ introduced 1D audio flow along the frequency axis for spectral morphing~\citep{ezzat2005audioflow}. No prior work computes standard 2D displacement fields on spectrogram images. Our framework fills this gap, directly capturing how spectral energy distributions shift across the time-frequency plane between consecutive frames.

\section{Datasets and Clinical Rationale}
\label{sec:data}

We evaluate on three corpora selected because their elicitation tasks recruit the multi-domain cognitive processes that underpin IADLs: planning, sequencing, monitoring, lexical retrieval, and sustained attention. Table~\ref{tab:datasets} summarizes the clinical scaffolding.

\begin{table}[ht]
\centering
\caption{Corpus characteristics. All three elicitation paradigms tax executive, attentional, and memory systems whose decline defines IADL impairment.}
\label{tab:datasets}
\begin{tabular}{lccc}
\toprule
& \textbf{Pitt (English)} & \textbf{EWA-DB (Slovak)} & \textbf{Ivanova (Spanish)} \\
\midrule
Speakers & $\sim$397 & 1,649 & 361 \\
Cognitive scale & MMSE + CDR & MoCA & MMSE \\
Core task & Cookie Theft PD & Naming + PD + DDK & Reading aloud \\
Recording period & 1983--2006 & 2020--2023 & 2018--2021 \\
Longitudinal & Yes & No & No \\
IADL link & CDR functional domains & MoCA executive items & MMSE attention items \\
\bottomrule
\multicolumn{4}{l}{\footnotesize PD = picture description; DDK = diadochokinesis.}
\end{tabular}
\end{table}

\textbf{English: DementiaBank Pitt Corpus.} The Cookie Theft picture description task from the Boston Diagnostic Aphasia Examination requires the speaker to construct a coherent narrative from a complex visual scene depicting a domestic kitchen. This taxes executive function (planning and sequencing the description), working memory (tracking what has been said), selective attention (prioritizing relevant scene elements), and lexical retrieval~\citep{cummings2019cookie,forbes2005detecting}. The scene itself depicts ADL activities (dishwashing, cooking), and Forbes-McKay and Venneri showed that over 70\% of patients with minimal AD fall below cut-off on semantic processing during this task, at precisely the disease stage where IADL impairment emerges~\citep{forbes2005detecting}. Clinical metadata includes MMSE scores and CDR values, the latter incorporating functional domains (Community Affairs, Home and Hobbies, Personal Care) that serve as an indirect bridge to IADL assessment~\citep{lanzi2023dementiabank}. This corpus, however, spans multiple decades of recording with heterogeneous equipment, a limitation we discuss in Section~\ref{sec:results}.

\textbf{Slovak: EWA-DB.} The largest corpus, with 1,649 speakers including 87 AD and 62 MCI participants assessed using the Montreal Cognitive Assessment~\citep{rusko2024slovak}. Its protocol is the most diverse: sustained vowel phonation, diadochokinetic sequences (/pataka/), 30-item object and action naming, and five picture descriptions. The naming and DDK tasks specifically probe articulatory planning and motor sequencing, cognitive demands that parallel the sequential planning required for IADLs such as medication management and meal preparation. All recordings were collected under standardized conditions with consistent equipment, eliminating the recording-environment confounds present in the Pitt Corpus.

\textbf{Spanish: Ivanova.} The Ivanova collection contains 361 European Spanish speakers (197 HC, 90 MCI, 74 AD) with MMSE scores as the primary cognitive measure~\citep{ivanova2022spanish}. The elicitation task is reading aloud the opening of \textit{Don Quixote}, a standardized passage that demands sustained attention, articulatory precision, and prosodic control. While lacking the spontaneous discourse demands of picture description, the reading task isolates speech production mechanisms from conceptual planning, enabling cleaner assessment of articulatory and phonatory integrity.

\textbf{Binary formulation.} MCI participants are merged into the AD class across all corpora, reflecting that MCI frequently represents a prodromal stage sharing overlapping speech biomarkers with early AD. This yields binary (AD vs.\ CN) classification consistent with ADReSS/ADReSSo evaluation protocols. Stratified speaker-level splits (65/15/20\% train/val/test) ensure no speaker appears in multiple partitions.

\section{Proposed Framework}
\label{sec:method}

The framework operates in two stages: a \textbf{segment model} that extracts and fuses acoustic and spectral dynamics features from short temporal windows, and a \textbf{speaker aggregator} that reasons over the sequence of segment representations to produce a patient-level decision.

\subsection{Preprocessing}

Each recording is resampled to 16\,kHz and converted to a log-power Mel spectrogram with 128 frequency bins, computed using a 1024-point FFT with 10\,ms hop and 25\,ms window. The spectrogram is segmented into 4-second overlapping windows yielding segments. During training, SpecAugment (frequency and time masking) and additive Gaussian noise provide regularization. Class imbalance is addressed through inverse-frequency weighted sampling.

\subsection{Acoustic Representation}

Each segment passes through a three-layer convolutional frontend that progressively builds spectral feature hierarchies, with batch normalization, ReLU activations, and max-pooling in the first two layers to compress the frequency dimension. The resulting feature maps are processed temporally through a Convolutional GRU that models sequential dependencies while preserving the 2D frequency structure of the representation. Unlike a standard GRU that flattens spatial dimensions, the ConvGRU replaces all linear transformations with 2D convolutions, enabling it to maintain frequency-localized hidden states. At each time step, the hidden state is globally average-pooled and projected to yield the acoustic embedding.

A critical implementation detail: the ConvGRU cell uses two separate convolutions rather than one. The first computes reset and update gates jointly; the second computes the candidate state from the input concatenated with the \textit{reset-gated} hidden state. If all three components are computed from a single convolution, the reset gate is calculated but never applied, effectively reducing the cell to a linear interpolation unit (see Appendix~\ref{app:gating} for details).

\subsection{Spectral Dynamics Representation}

The core premise is that dementia-related vocal changes, such as formant trajectory deviations, pitch contour instability, and irregular energy redistribution, manifest as systematic shifts in how spectral energy is distributed across the time-frequency plane between consecutive frames. In healthy speech, these frame-to-frame transitions follow regular, periodic patterns reflecting stable articulatory control. In impaired speech, the transitions become erratic.

To capture this, consecutive spectrogram frames are stacked along the channel dimension and passed through a convolutional encoder that computes dense displacement fields. The encoder uses progressively smaller kernels to capture spectral changes at multiple scales: large receptive fields detect broad energy redistribution patterns associated with speech rate variation, while smaller kernels capture fine-grained perturbations such as cycle-to-cycle formant jitter. The output is a 2-channel field that represents the spectral displacement at every point in the time-frequency plane. Trajectories are transferred from this field and encoded through a two-layer MLP to produce the spectral dynamics embedding.

\subsection{Cross-Attention Fusion}

The two representations encode complementary information: the acoustic embedding captures \textit{what} the spectral content is at each moment, while the spectral dynamics embedding captures \textit{how} that content evolves. To integrate them without destructive interference, the spectral dynamics vector serves as query while the acoustic temporal sequence provides keys and values:
\begin{equation}
    \mathbf{f}_i = \text{LayerNorm}\!\left(\bQ + \text{MultiHead}(\bQ, \bK, \bV)\right), \quad \bQ = \bM_i^{\top}, \;\; \bK{=}\bV{=}\bA_i
    \label{eq:crossattn}
\end{equation}
with 4 attention heads and embedding dimension 128. This allows the spectral dynamics signature to selectively attend to acoustic time intervals where spectrotemporal anomalies co-occur with vocal degradation. A linear projection maps the fused feature to the model dimension.

\subsection{Temporal Aggregation}

A recording typically produces 5--30 overlapping segments. These are embedded with sinusoidal positional encodings and processed through a two-layer Transformer encoder (4 heads, FFN dim 512) that enables each segment to attend to all others, learning long-range temporal dependencies such as fatigue effects or progressive disfluency. A learnable query vector then attends to the contextualized sequence to produce a single patient representation $\bp$, which is classified by a two-layer MLP.

\subsection{Composite Temporal Loss}

The total objective combines cross-entropy (with label smoothing $\epsilon{=}0.1$) with five auxiliary terms that regularize the temporal structure of segment-level predictions:
\begin{equation}
    \cL = \cL_{\text{CE}} + \lambda_{\text{TC}}\cL_{\text{TC}} + \lambda_{\text{CL}}\cL_{\text{CL}} + \lambda_{\text{P}}\cL_{\text{P}} + \lambda_{\text{CH}}\cL_{\text{CH}} + \lambda_{\text{AE}}\cL_{\text{AE}}
    \label{eq:loss}
\end{equation}
Temporal consistency ($\lambda{=}0.01$) penalizes abrupt probability shifts between consecutive segments. Temporal contrastive ($\lambda{=}0.05$) applies InfoNCE with temperature 0.07 to encourage adjacent segments to be similar in feature space. Progression ($\lambda{=}0.05$) penalizes second-order oscillations in the AD probability trajectory. Multi-scale coherence ($\lambda{=}0.01$) enforces consistency across temporal resolutions via average-pooling at scales 2 and 4. Attention entropy ($\lambda{=}0.01$) encourages focused pooling. Full definitions are in Appendix~\ref{app:losses}.

\subsection{Training}

All models use AdamW (lr\,=\,$2{\times}10^{-4}$, weight decay $10^{-4}$), cosine annealing warm restarts, gradient accumulation over 4 steps (effective batch 64), gradient clipping at 1.0, and mixed-precision training. Each language is trained independently with identical hyperparameters, yielding three separate models. Four ablation configurations (full, audio-only, spectral-dynamics-only, no-attention) are evaluated on both Spanish and English. The best checkpoint is selected by validation AUC + accuracy.

\section{Experimental Results}
\label{sec:results}

\subsection{Main Results}

\begin{table}[ht]
\centering
\caption{Test set performance (full model). English$^\dagger$ serves as baseline.}
\label{tab:main}
\begin{tabular}{lcccccc}
\toprule
\textbf{Language} & \textbf{Acc.} & \textbf{AUC} & \textbf{F1} & \textbf{Prec.} & \textbf{Rec.} & \textbf{Loss} \\
\midrule
English$^\dagger$ & 0.532 & 0.563 & 0.522 & 0.522 & 0.532 & 0.789 \\
Slovak & \textbf{0.839} & 0.755 & \textbf{0.878} & \textbf{0.937} & \textbf{0.839} & \textbf{0.617} \\
Spanish & 0.685 & \textbf{0.788} & 0.663 & 0.708 & 0.685 & 0.705 \\
\bottomrule
\multicolumn{7}{l}{\footnotesize $^\dagger$DementiaBank Pitt Corpus.}
\end{tabular}
\end{table}

The Slovak model achieves 83.9\% accuracy and 93.7\% precision, indicating confident dementia identification with minimal false positives. The Spanish model achieves the highest AUC (0.788), indicating superior discriminative ranking despite lower accuracy, likely because the fixed threshold of 0.5 is suboptimal for its class distribution. The English baseline yields 53.2\% accuracy (AUC 0.563), consistent with Liu et al.'s findings that Pitt Corpus discriminability partly reflects recording artifacts~\citep{liu2024cleverhans}. The multi-decade recording heterogeneity and absence of speaker diarization (interviewer speech contaminates the signal) are the most probable causes.

\begin{figure}[ht]
\centering
\begin{subfigure}[b]{0.32\textwidth}
    \includegraphics[width=\textwidth]{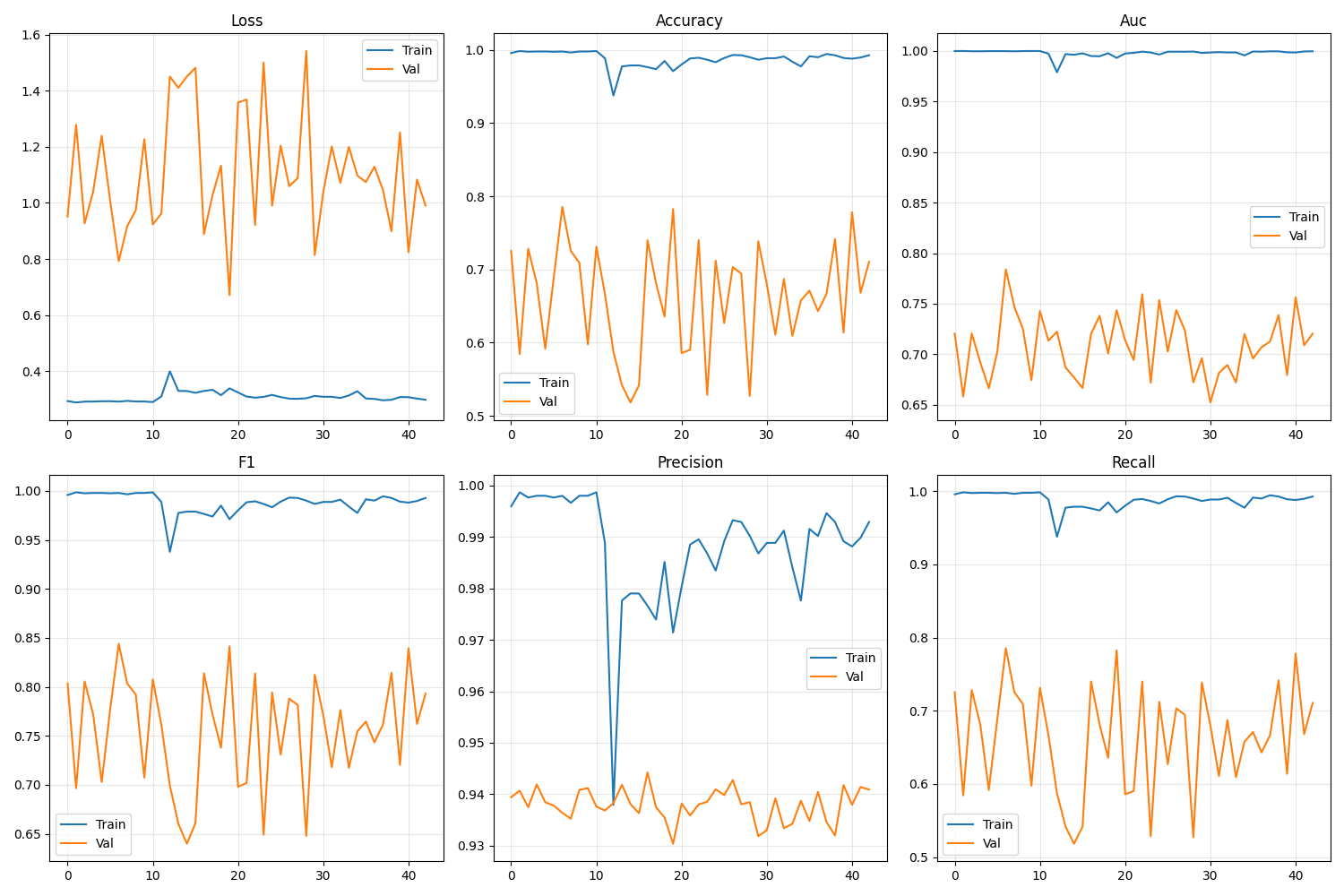}
    \caption{Slovak: training curves}
\end{subfigure}\hfill
\begin{subfigure}[b]{0.32\textwidth}
    \includegraphics[width=\textwidth]{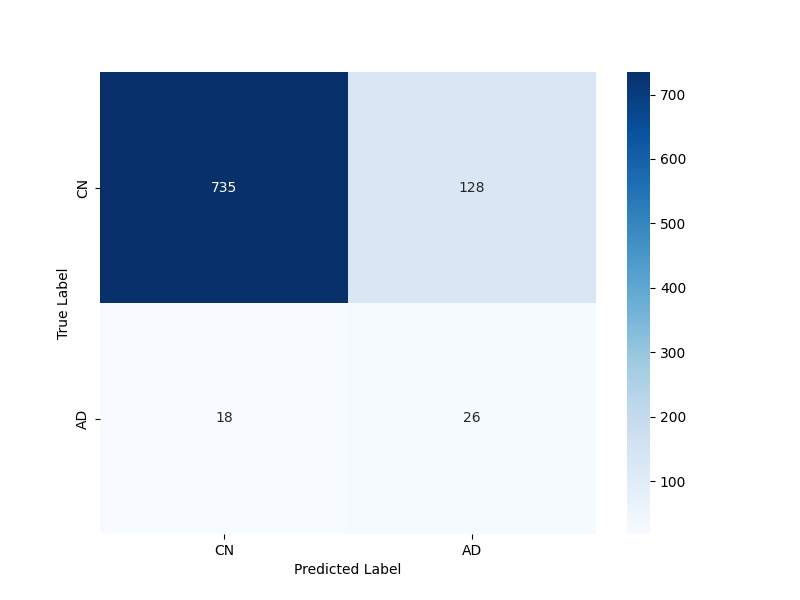}
    \caption{Slovak: confusion matrix}
\end{subfigure}\hfill
\begin{subfigure}[b]{0.32\textwidth}
    \includegraphics[width=\textwidth]{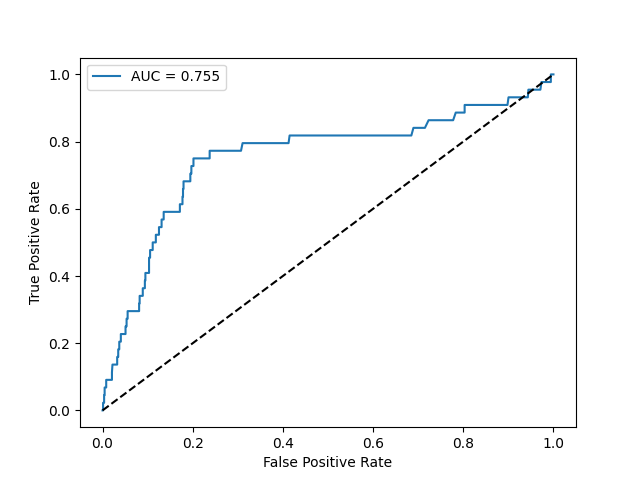}
    \caption{Slovak: ROC curve}
\end{subfigure}
\caption{Slovak corpus, full model. Training converges to near-perfect performance while validation stabilizes}
\label{fig:slovak}
\end{figure}

\begin{figure}[ht]
\centering
\begin{subfigure}[b]{0.32\textwidth}
    \includegraphics[width=\textwidth]{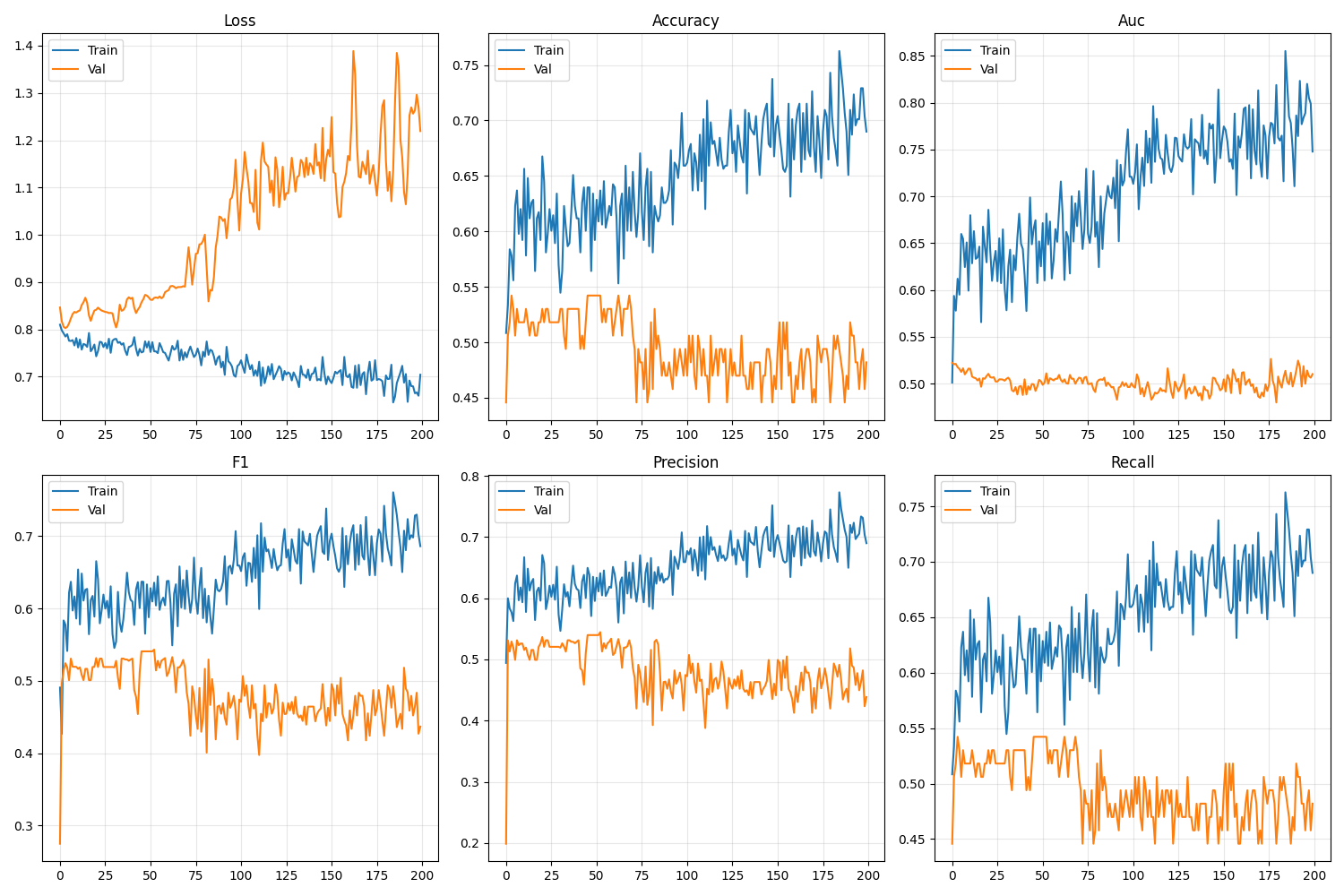}
    \caption{English: training curves}
\end{subfigure}\hfill
\begin{subfigure}[b]{0.32\textwidth}
    \includegraphics[width=\textwidth]{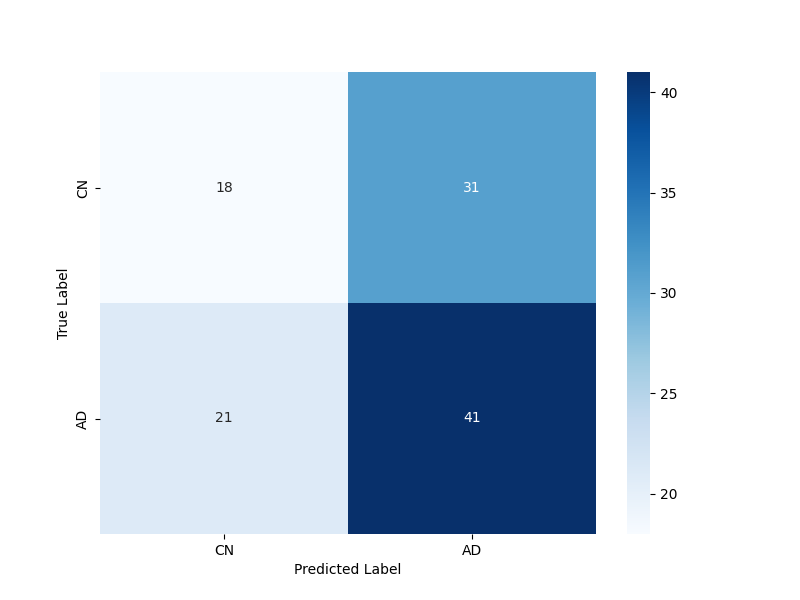}
    \caption{English: confusion matrix}
\end{subfigure}\hfill
\begin{subfigure}[b]{0.32\textwidth}
    \includegraphics[width=\textwidth]{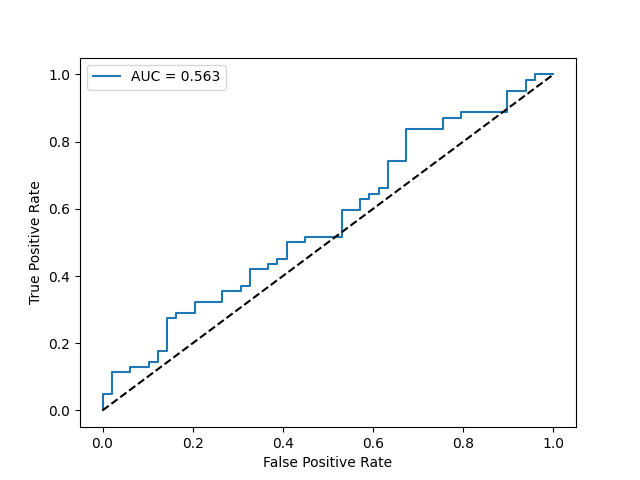}
    \caption{English: ROC curve}
\end{subfigure}
\caption{English baseline, full model. Training loss decreases while validation loss increases sharply, indicating memorization of speaker-specific characteristics.}
\label{fig:english}
\end{figure}

\subsection{Auxiliary Loss Invariance}

\begin{table}[ht]
\centering
\caption{Converged loss decomposition. Auxiliary losses are stable across languages (CV\,$\leq$\,7\%), including the  English baseline, confirming performance-invariant regularization.}
\label{tab:loss}
\begin{tabular}{lccccc}
\toprule
\textbf{Component} & \textbf{English} & \textbf{Slovak} & \textbf{Spanish} & \textbf{Mean} & \textbf{CV (\%)} \\
\midrule
$\cL_{\text{CE}}$ & 0.550 & 0.204 & 0.714 & 0.489 & 53.6 \\
$\cL_{\text{TC}}$ & $2.8{\times}10^{-6}$ & $6.1{\times}10^{-9}$ & $2.1{\times}10^{-5}$ & $\approx$0 & -- \\
$\cL_{\text{feat}}$ & 0.060 & 0.056 & 0.058 & 0.058 & 3.4 \\
$\cL_{\text{AE}}$ & 0.034 & 0.031 & 0.030 & 0.032 & 6.7 \\
\midrule
\textbf{Total} & 0.645 & 0.292 & 0.802 & 0.580 & 44.5 \\
\bottomrule
\end{tabular}
\end{table}

Cross-entropy varies dramatically across languages (CV\,=\,53.6\%), reflecting differences in classification difficulty. In contrast, the feature losses cluster tightly (CV\,=\,3.4\%) and the attention entropy converges within 6.7\% across all three languages, including English where classification fails. This performance-invariant convergence confirms that the temporal regularization framework imposes structurally valid constraints independent of language or primary task success.

\subsection{Generalization Regimes}

\begin{table}[ht]
\centering
\caption{Train / validation / test comparison.}
\label{tab:gen}
\begin{tabular}{llcccccc}
\toprule
\textbf{Lang.} & \textbf{Split} & \textbf{Acc.} & \textbf{AUC} & \textbf{F1} & \textbf{Prec.} & \textbf{Rec.} & \textbf{Loss} \\
\midrule
\multirow{3}{*}{English} & Train & 0.690 & 0.748 & 0.686 & 0.690 & 0.690 & 0.704 \\
& Val & 0.482 & 0.510 & 0.437 & 0.438 & 0.482 & 1.219 \\
& \textbf{Test} & \textbf{0.532} & \textbf{0.563} & \textbf{0.522} & \textbf{0.522} & \textbf{0.532} & \textbf{0.789} \\
\midrule
\multirow{3}{*}{Slovak} & Train & 0.993 & 1.000 & 0.993 & 0.993 & 0.993 & 0.298 \\
& Val & 0.711 & 0.720 & 0.793 & 0.941 & 0.711 & 0.990 \\
& \textbf{Test} & \textbf{0.839} & \textbf{0.755} & \textbf{0.878} & \textbf{0.937} & \textbf{0.839} & \textbf{0.617} \\
\midrule
\multirow{3}{*}{Spanish} & Train & 0.682 & 0.710 & 0.679 & 0.692 & 0.682 & 0.741 \\
& Val & 0.639 & 0.791 & 0.626 & 0.638 & 0.639 & 0.713 \\
& \textbf{Test} & \textbf{0.685} & \textbf{0.788} & \textbf{0.663} & \textbf{0.708} & \textbf{0.685} & \textbf{0.705} \\
\bottomrule
\end{tabular}
\end{table}

Three regimes emerge. English shows generalization failure: validation falls below chance (48.2\%), indicating memorization of speaker identity rather than cognitive biomarkers. Slovak exhibits productive overfitting: near-perfect training but test exceeds validation by 12.8pp, indicating successful early stopping. Spanish shows data-limited behavior: negligible train-test gap ($-$0.3pp), with validation AUC exceeding training AUC, indicating implicit regularization from augmentation.

\subsection{Ablation: Spanish}

\begin{table}[ht]
\centering
\caption{Spanish ablation. Removing cross-attention collapses performance below either unimodal model.}
\label{tab:abl_es}
\begin{tabular}{lcccccc}
\toprule
\textbf{Config.} & \textbf{Acc.} & \textbf{AUC} & \textbf{F1} & \textbf{Prec.} & \textbf{Rec.} & \textbf{Loss} \\
\midrule
Full model & 0.685 & \textbf{0.788} & 0.663 & 0.708 & 0.685 & 0.705 \\
Audio only & 0.685 & 0.659 & 0.655 & \textbf{0.725} & 0.685 & 0.718 \\
Spectral dyn. & 0.685 & 0.603 & 0.663 & 0.708 & 0.685 & 0.710 \\
No attention & 0.537 & 0.567 & 0.514 & 0.521 & 0.537 & 0.782 \\
\bottomrule
\end{tabular}
\end{table}

\begin{figure}[ht]
\centering
\begin{subfigure}[b]{0.32\textwidth}
    \includegraphics[width=\textwidth]{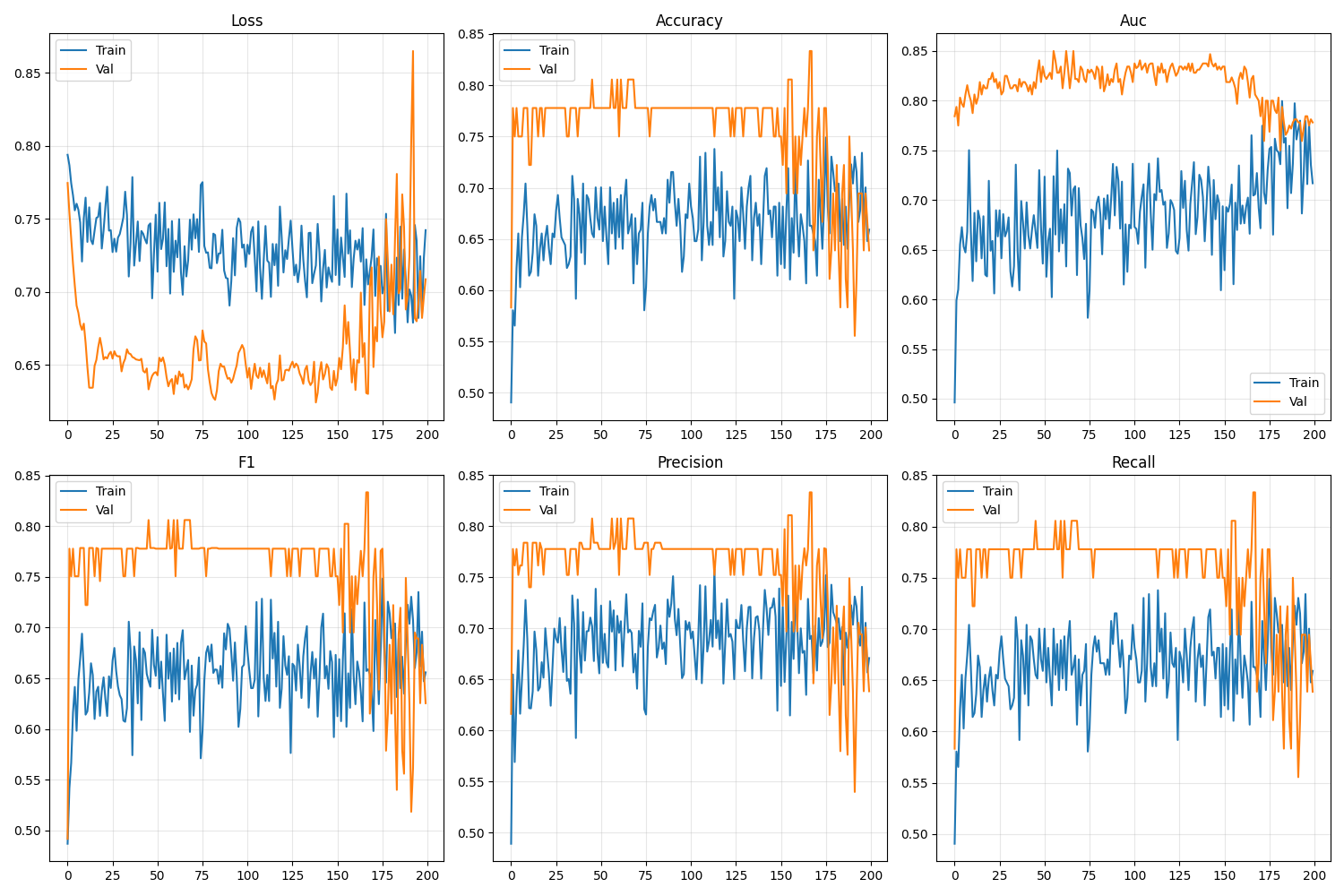}
    \caption{Training curves}
\end{subfigure}\hfill
\begin{subfigure}[b]{0.32\textwidth}
    \includegraphics[width=\textwidth]{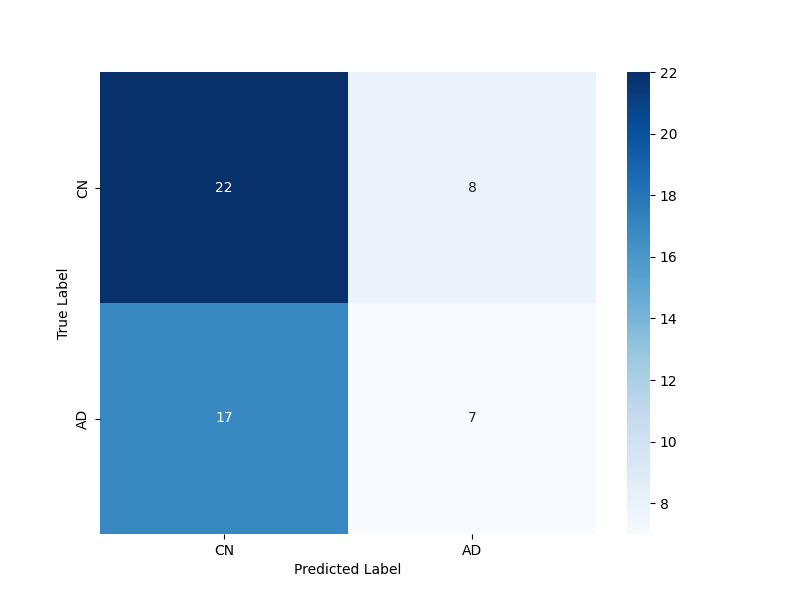}
    \caption{Confusion matrix}
\end{subfigure}\hfill
\begin{subfigure}[b]{0.32\textwidth}
    \includegraphics[width=\textwidth]{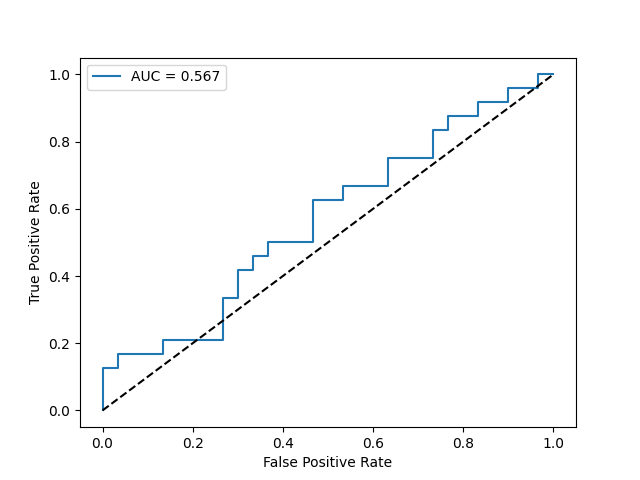}
    \caption{ROC curve}
\end{subfigure}
\caption{Spanish no-attention ablation. Naive element-wise addition of the two modalities produces catastrophic collapse to near-chance performance, demonstrating destructive interference.}
\label{fig:noattn}
\end{figure}

Three findings emerge from the Spanish ablation. First, the full model, audio-only, and spectral-dynamics-only configurations achieve identical accuracy (68.5\%), yet their AUCs diverge sharply (0.788, 0.659, 0.603). The benefit of multimodal fusion lies in improved rank-ordering, not point-estimate accuracy at a fixed threshold. Second, removing cross-attention while retaining both modalities drops accuracy to 53.7\%, below either unimodal model. This is a clinical instantiation of destructive modality competition: naive element-wise addition allows noise from one branch to corrupt signal from the other. Third, the full model's AUC (0.788) exceeds both unimodal AUCs confirming synergistic rather than redundant information fusion.

\subsection{Ablation: Slovak}

\begin{table}[ht]
\centering
\caption{Slovak ablation. Audio-only \textit{outperforms} the full model. Spectral dynamics alone achieves 95.1\% accuracy but near-chance AUC (0.506), indicating majority-class prediction.}
\label{tab:abl_sk}
\begin{tabular}{lcccccc}
\toprule
\textbf{Config.} & \textbf{Acc.} & \textbf{AUC} & \textbf{F1} & \textbf{Prec.} & \textbf{Rec.} & \textbf{Loss} \\
\midrule
Full model & 0.839 & 0.755 & 0.878 & 0.937 & 0.839 & 0.617 \\
Audio only & \textbf{0.937} & \textbf{0.766} & \textbf{0.932} & 0.928 & \textbf{0.937} & \textbf{0.408} \\
Spectral dyn. & 0.951 & 0.506 & 0.928 & 0.905 & 0.951 & 0.737 \\
\bottomrule
\end{tabular}
\end{table}

The Slovak ablation reveals a pattern that \textit{inverts} the Spanish findings. Audio-only achieves 93.7\% accuracy and 0.766 AUC, both \textit{exceeding} the full model (83.9\%, 0.755). This indicates that on a standardized, acoustically homogeneous corpus, the CNN-ConvGRU acoustic encoder captures sufficient discriminative signal on its own, and the spectral dynamics branch introduces noise that the cross-attention mechanism only partially filters.

The spectral-dynamics-only configuration exposes this most clearly. It achieves the highest raw accuracy (95.1\%) but near-chance AUC (0.506), a pathological dissociation indicating that the model has learned to predict the majority class with high confidence rather than discriminating between classes. The 95.1\% accuracy reflects the class imbalance in the test set rather than genuine classification ability.

This suggests that on a high-quality corpus where acoustic features alone are sufficient, the attention mechanism may over-align the two modalities, introducing unnecessary complexity that hinders optimization.

\subsection{Ablation: English}

\begin{table}[ht]
\centering
\caption{English ablation. All configurations cluster near chance, confirming corpus-level limitations.}
\label{tab:abl_en}
\begin{tabular}{lcccccc}
\toprule
\textbf{Config.} & \textbf{Acc.} & \textbf{AUC} & \textbf{F1} & \textbf{Prec.} & \textbf{Rec.} & \textbf{Loss} \\
\midrule
Full model & 0.532 & 0.563 & 0.522 & 0.522 & 0.532 & 0.789 \\
Audio only & 0.505 & 0.537 & 0.505 & 0.516 & 0.505 & 0.818 \\
Spectral dyn. & \textbf{0.559} & 0.549 & 0.400 & 0.312 & \textbf{0.559} & 0.801 \\
No attention & 0.541 & \textbf{0.573} & \textbf{0.533} & \textbf{0.532} & 0.541 & 0.791 \\
\bottomrule
\end{tabular}
\end{table}

The English ablation reveals a fundamentally different pattern. All four configurations cluster in a narrow band (accuracy 50.5--55.9\%, AUC 0.537--0.573), and the no-attention configuration does \textit{not} collapse but performs comparably to the full model. This confirms that when the underlying corpus lacks genuine discriminative signal, no architectural variation can compensate.

\begin{table}[ht]
\centering
\caption{Component contribution ($\Delta$ vs.\ full model) across all three languages. Each language exhibits a distinct fusion regime.}
\label{tab:delta}
\begin{tabular}{l cc cc cc}
\toprule
& \multicolumn{2}{c}{\textbf{Slovak}} & \multicolumn{2}{c}{\textbf{Spanish}} & \multicolumn{2}{c}{\textbf{English}} \\
\cmidrule(lr){2-3} \cmidrule(lr){4-5} \cmidrule(lr){6-7}
\textbf{Removed} & $\Delta$Acc & $\Delta$AUC & $\Delta$Acc & $\Delta$AUC & $\Delta$Acc & $\Delta$AUC \\
\midrule
Spectral dyn. & +9.8pp & +0.011 & 0.0 & $-$0.129 & $-$2.7pp & $-$0.026 \\
Audio encoder & +11.2pp & $-$0.249 & 0.0 & $-$0.185 & +2.7pp & $-$0.014 \\
Cross-attention & $-$2.2pp & \textbf{+0.117} & \textbf{$-$14.8pp} & \textbf{$-$0.221} & +0.9pp & +0.010 \\
\bottomrule
\end{tabular}
\end{table}

Table~\ref{tab:delta} reveals three distinct fusion regimes. On Spanish, cross-attention is essential: removing it causes catastrophic collapse , confirming that learned alignment is necessary when neither modality alone captures sufficient signal. On Slovak, cross-attention is unnecessary: the audio encoder alone outperforms the full model (+9.8pp), and removing attention actually improves AUC, indicating that on a high-quality corpus the acoustic signal is self-sufficient and multimodal fusion introduces optimization overhead. In English, all deltas are negligible ($<$3pp), which confirms the behavior of the noise-floor where no architectural variation can compensate for corpus-level limitations. The value of the fusion mechanism is corpus-dependent: it is critical when the signal is distributed across modalities (Spanish), counterproductive when one modality dominates (Slovak), and irrelevant when no signal exists (English).

\subsection{Feature Space Visualization}

\begin{figure}[ht]
\centering
\begin{subfigure}[b]{0.48\textwidth}
    \includegraphics[width=\textwidth]{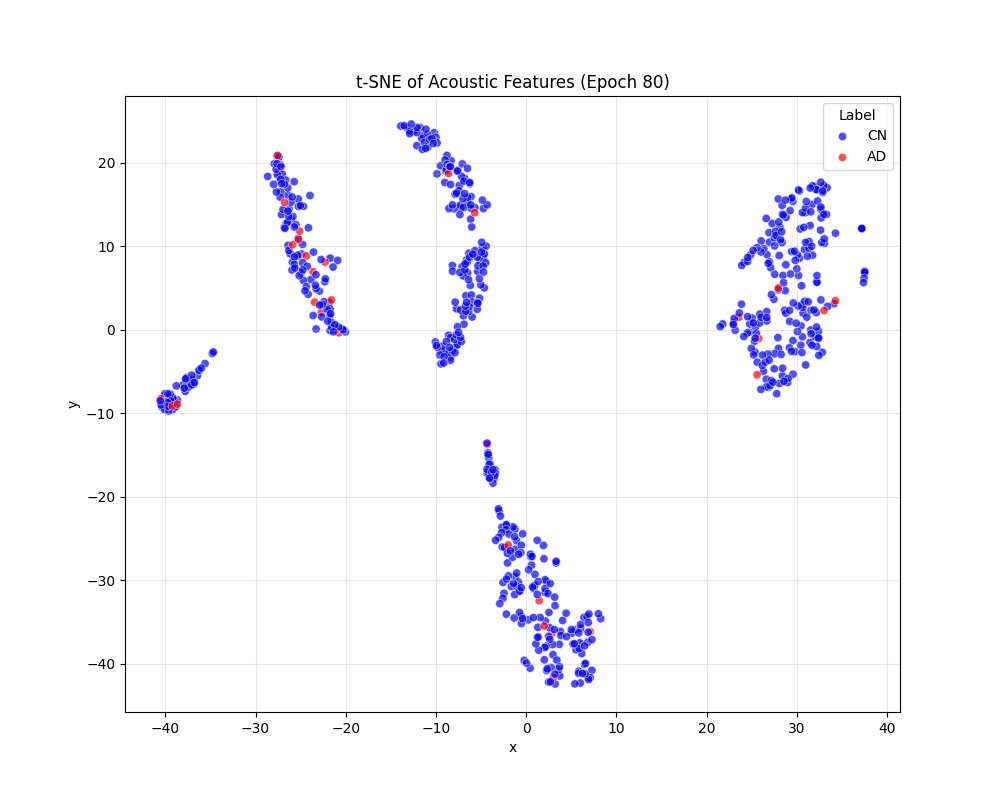}
    \caption{Epoch 80}
\end{subfigure}\hfill
\begin{subfigure}[b]{0.48\textwidth}
    \includegraphics[width=\textwidth]{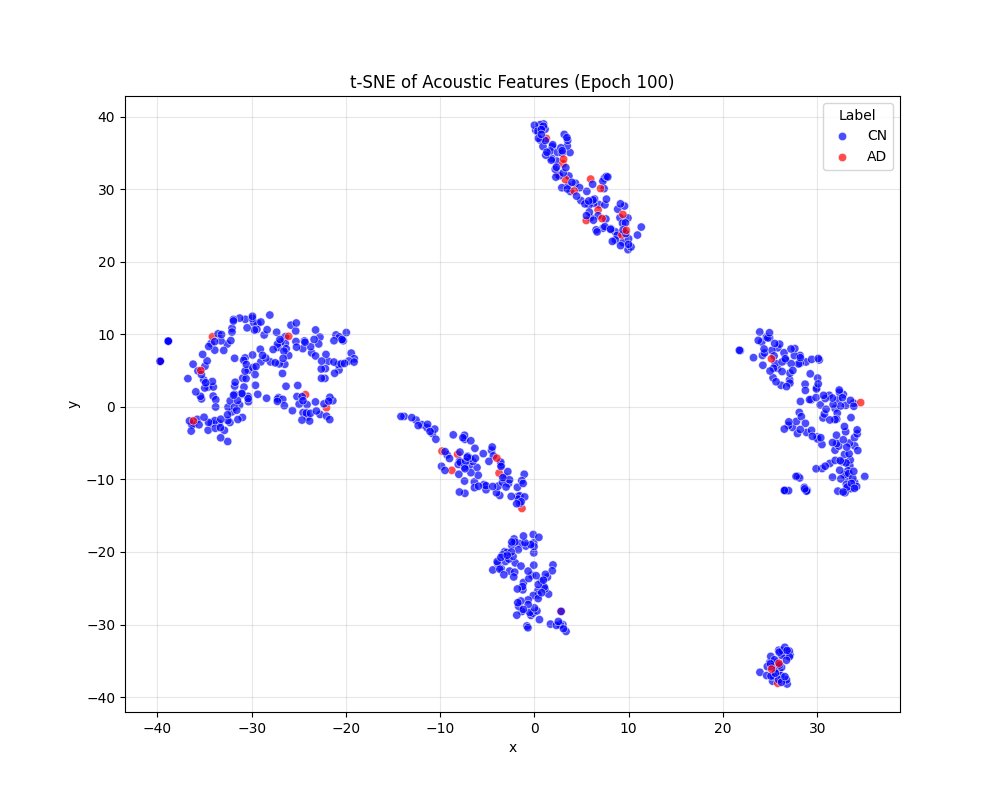}
    \caption{Epoch 100}
\end{subfigure}
\caption{t-SNE of acoustic features on the Slovak validation set. The model forms distinct clusters, but AD samples (red) remain distributed across CN-dominated clusters rather than forming a separable region, consistent with the observed generalization gap.}
\label{fig:tsne}
\end{figure}

\section{Discussion}
\label{sec:discussion}

\textbf{Three fusion regimes emerge across corpora.} The complete ablation across all three languages reveals that cross-attention fusion is not universally beneficial but rather corpus-dependent. On Spanish, it is essential: removing it causes catastrophic collapse below either unimodal baseline. On Slovak, it is counterproductive: the audio encoder alone outperforms the full model by accuracy, and removing attention improves AUC by 0.117. On English, it is irrelevant: all configurations cluster near chance. This three-regime pattern suggests that cross-attention's value depends on the \textit{distribution of discriminative signal} across modalities. When signal is distributed (Spanish), learned alignment is necessary to combine complementary information without destructive interference. When signal is concentrated in one modality (Slovak), the attention mechanism introduces optimization overhead that hinders the dominant branch.

\textbf{Corpus quality remains the dominant factor.} The identical architecture yields 93.7\% accuracy (audio-only) on standardized Slovak recordings and 53.2\% on heterogeneous Pitt Corpus recordings. The English ablation confirms this: all four configurations produce near-chance results. The field must prioritize corpus standardization, speaker diarization, and acoustic normalization before architectural innovation can be meaningfully evaluated.

\textbf{The spectral dynamics encoder reveals a class-imbalance artifact on Slovak.} The spectral-dynamics-only Slovak model achieves 95.1\% accuracy but 0.506 AUC, a pathological dissociation indicating majority-class prediction rather than genuine discrimination. On Spanish, the same encoder achieves balanced accuracy and AUC (0.685/0.603). This contrast suggests that the displacement field representation is more susceptible to class-imbalance exploitation on larger, more imbalanced corpora, and that AUC should be the primary metric for evaluating spectral dynamics contributions.

\textbf{Speech as an IADL proxy.} The performance hierarchy aligns with the quality and cognitive complexity of each corpus's clinical protocol rather than with language properties. The Slovak corpus, with its diverse protocol taxing multiple IADL-relevant cognitive domains under controlled conditions, produces the best unimodal results. The Spanish reading task, which isolates speech production from conceptual planning, shows the greatest benefit from multimodal fusion. This supports the hypothesis that speech assessment is most effective when elicitation tasks recruit executive, attentional, and planning resources.

\textbf{Limitations.} (1) Speaker diarization is not applied, particularly impacting English. (2) Corpus sizes differ across languages. 

\section{Conclusion}
\label{sec:conclusion}

We introduced an ASR-agnostic framework that extracts spectrotemporal displacement fields from speech spectrograms, fuses them with CNN-ConvGRU acoustic features through learned cross-attention, and enforces temporal coherence through composite regularization. Complete ablation across three languages reveals three distinct fusion regimes: cross-attention is essential when discriminative signal is distributed across modalities (Spanish), counterproductive when acoustic features dominate (Slovak), and irrelevant when corpus-level artifacts preclude genuine classification (English). Temporal regularization operates in a language- and performance-invariant manner across all regimes. These findings argue that the value of multimodal fusion in clinical speech analysis depends fundamentally on the interaction between corpus quality and signal distribution, and that no single architectural choice is universally optimal. The alignment between our strongest results and the most cognitively demanding elicitation protocols supports the broader premise that speech analysis, when grounded in tasks that tax IADL-relevant cognitive processes, can serve as a scalable proxy for functional cognitive assessment.

\newpage
\appendix
\section{Mathematical Details}
\label{app:math}

\subsection{Notation}

\begin{table}[ht]
\centering
\caption{Notation summary.}
\label{tab:notation}
\begin{tabular}{ll}
\toprule
\textbf{Symbol} & \textbf{Description} \\
\midrule
$\mathbf{S} \in \R^{128 \times T}$ & Log-Mel spectrogram (128 bins, $T$ frames at 10\,ms) \\
$\mathbf{S}_i \in \R^{128 \times 400}$ & $i$-th segment (4 seconds, 50\% overlap) \\
$\bA_i \in \R^{T' \times 128}$ & Audio encoder output \\
$\bM_i \in \R^{128}$ & Spectral dynamics encoder output \\
$\mathbf{f}_i \in \R^{256}$ & Fused segment feature \\
$\boldsymbol{\Phi} \in \R^{2 \times H \times W}$ & Spectral displacement field \\
$\bC \in \R^{N \times 256}$ & Contextualized sequence \\
$\bp \in \R^{256}$ & Pooled patient representation \\
\bottomrule
\end{tabular}
\end{table}

\subsection{ConvGRU Gate Equations}
\label{app:gating}

The ConvGRU uses two separate convolutions. The gate convolution takes $[\bx_t, \bh_{t-1}]$ and outputs $2C_h$ channels split into update gate $\bz_t$ and reset gate $\br_t$:
\begin{align}
    \bz_t &= \sigma(\bW_z * [\bx_t, \bh_{t-1}]), \quad
    \br_t = \sigma(\bW_r * [\bx_t, \bh_{t-1}])
\end{align}
The candidate convolution takes $[\bx_t, \br_t \odot \bh_{t-1}]$:
\begin{align}
    \tilde{\bh}_t &= \tanh(\bW_h * [\bx_t, \br_t \odot \bh_{t-1}]), \quad
    \bh_t = (1 - \bz_t) \odot \bh_{t-1} + \bz_t \odot \tilde{\bh}_t
\end{align}
where $*$ denotes 2D convolution (kernel $3{\times}3$, 128 channels) and $\odot$ is the Hadamard product.

\subsection{Multi-Head Attention}
\begin{align}
    \text{head}_j &= \text{softmax}\!\left(\frac{(\bQ \bW_j^Q)(\bK \bW_j^K)^\top}{\sqrt{d_k}}\right) \bV \bW_j^V, \quad d_k = 32 \\
    \text{MultiHead} &= \text{Concat}(\text{head}_1, \ldots, \text{head}_4)\bW^O
\end{align}

\subsection{Temporal Loss Definitions}
\label{app:losses}

\textbf{Temporal consistency.} For segment probabilities $\bp_t = \text{softmax}(\text{logits}_t)$:
\begin{equation}
    \cL_{\text{TC}} = \frac{1}{N-1}\sum_{t=1}^{N-1}\|\bp_{t+1} - \bp_t\|_2
\end{equation}

\textbf{Temporal contrastive (InfoNCE).} For L2-normalized features $\{\mathbf{g}_t\}$ with neighbor set $\mathcal{N}(t) = \{t{-}1, t{+}1\}$:
\begin{equation}
    \cL_{\text{CL}} = -\frac{1}{N}\sum_{t=1}^{N}\log\frac{\sum_{j \in \mathcal{N}(t)} \exp(\mathbf{g}_t^\top \mathbf{g}_j / \tau)}{\sum_{j \in \mathcal{N}(t)} \exp(\mathbf{g}_t^\top \mathbf{g}_j / \tau) + \sum_{k \notin \mathcal{N}(t) \cup \{t\}} \exp(\mathbf{g}_t^\top \mathbf{g}_k / \tau)}
\end{equation}

\textbf{Progression.} Second-order smoothness on AD probability $p_t = \text{softmax}(\text{logits}_t)_{[1]}$:
\begin{equation}
    \cL_{\text{P}} = \frac{1}{N-2}\sum_{t=1}^{N-2}|\Delta^2 p_t|
\end{equation}

\subsection{Algorithm: Inference Pipeline}

\begin{algorithm}[H]
\caption{ASR-Agnostic Dementia Classification}
\label{alg:inference}
\SetAlgoLined
\KwIn{Waveform $\bx \in \R^{T_{\text{raw}}}$}
\KwOut{$\hat{y} \in \{0,1\}$, probability $p$}
$\mathbf{S} \gets \textsc{LogMelSpec}(\textsc{Resample}(\bx, 16\text{kHz}))$\;
$\{\mathbf{S}_i\}_{i=1}^{N} \gets \textsc{Segment}(\mathbf{S}; 4\text{s}, 50\%)$\;
\For{$i = 1, \ldots, N$}{
    $\bA_i \gets \textsc{AudioEnc}(\mathbf{S}_i)$\;
    $\boldsymbol{\Phi}_i \gets \textsc{DispNet}(\textsc{FramePair}(\mathbf{S}_i))$\;
    $\bM_i \gets \textsc{DynEnc}(\textsc{SampleTraj}(\boldsymbol{\Phi}_i))$\;
    $\mathbf{f}_i \gets \textsc{Bridge}(\textsc{CrossAttn}(\bM_i, \bA_i))$\;
}
$\bC \gets \textsc{TransformerEnc}(\textsc{PE}([\mathbf{f}_1, \ldots, \mathbf{f}_N]))$\;
$\bp \gets \textsc{QueryAttn}(\bq, \bC)$; \quad $p \gets \text{softmax}(\textsc{MLP}(\bp))_{[1]}$\;
$\hat{y} \gets \mathbb{1}[p > 0.5]$\;
\end{algorithm}

\subsection{Hyperparameters}

\begin{table}[ht]
\centering
\caption{Complete configuration (identical across all languages).}
\label{tab:hyp}
\begin{tabular}{lclc}
\toprule
\textbf{Param.} & \textbf{Value} & \textbf{Param.} & \textbf{Value} \\
\midrule
Sample rate & 16\,kHz & Mel bins & 128 \\
FFT / hop / win & 1024 / 160 / 400 & Segment & 4\,s, 50\% \\
$d_{\text{model}}$ & 256 & Encoder dim & 128 \\
Attn heads & 4 & TF layers & 2 \\
FFN dim & 512 & Dropout & 0.1 \\
Eff.\ batch & 64 & lr & $2{\times}10^{-4}$ \\
Weight decay & $10^{-4}$ & Grad clip & 1.0 \\
Epochs / patience & 50 / 15 & $\epsilon_{\text{smooth}}$ & 0.1 \\
$\lambda_{\text{TC,CH,AE}}$ & 0.01 & $\lambda_{\text{CL,P}}$ & 0.05 \\
$\tau$ (contrast) & 0.07 & Threshold & 0.5 \\
\bottomrule
\end{tabular}
\end{table}

\bibliographystyle{plainnat}

\begin{thebibliography}{99}

\bibitem[Agbavor \& Liang(2023)]{agbavor2023endtoend}
Agbavor, F. and Liang, H. (2023). AI-enabled end-to-end detection and assessment of Alzheimer's disease using voice. \textit{Brain Sciences}, 13(1):28.

\bibitem[Ahmed et al.(2013)]{ahmed2013connected}
Ahmed, S., et al. (2013). Connected speech as a marker of disease progression in autopsy-proven Alzheimer's disease. \textit{Brain}, 136(12):3727--3737.

\bibitem[Cho et al.(2022)]{cho2022prosodic}
Cho, S., et al. (2022). Prosodic characteristics of prepausal words produced by patients with neurodegenerative disease. \textit{Speech Prosody}.

\bibitem[Chu et al.(2023)]{chu2023audiovisual}
Chu, C.-S., et al. (2023). Automated video analysis of audio-visual approaches to predict and detect MCI and dementia. \textit{J. Alzheimer's Dis.}, 93(2).

\bibitem[Cummings(2019)]{cummings2019cookie}
Cummings, L. (2019). Describing the Cookie Theft picture: Sources of breakdown in Alzheimer's dementia. \textit{Pragmatics \& Society}, 10(2):151--174.

\bibitem[de la Fuente Garcia et al.(2020)]{delafuente2020systematic}
de la Fuente Garcia, S., Ritchie, C.W., and Luz, S. (2020). AI, speech, and language processing approaches to monitoring Alzheimer's disease: A systematic review. \textit{J. Alzheimer's Dis.}, 78(4):1547--1574.

\bibitem[Du et al.(2023)]{du2023unimodal}
Du, Y., et al. (2023). On uni-modal feature learning in supervised multi-modal learning. \textit{Proc. ICML}, PMLR 202.

\bibitem[Ezzat et al.(2005)]{ezzat2005audioflow}
Ezzat, T., et al. (2005). Morphing spectral envelopes using audio flow. \textit{Proc. Interspeech}.

\bibitem[Forbes-McKay \& Venneri(2005)]{forbes2005detecting}
Forbes-McKay, K.E. and Venneri, A. (2005). Detecting subtle spontaneous language decline in early Alzheimer's disease with a picture description task. \textit{Neurological Sciences}, 26(4):243--254.

\bibitem[Gao et al.(2025)]{gao2025dstcnet}
Gao, Y., et al. (2025). A dual-stage time-context network for speech-based AD detection. \textit{arXiv:2502.13064}.

\bibitem[Gong et al.(2021)]{gong2021ast}
Gong, Y., et al. (2021). AST: Audio Spectrogram Transformer. \textit{Proc. Interspeech}.

\bibitem[Haider et al.(2020)]{haider2020paralinguistic}
Haider, F., et al. (2020). An assessment of paralinguistic acoustic features for detection of AD in spontaneous speech. \textit{IEEE J. Sel. Topics Signal Process.}, 14(2).

\bibitem[Huang et al.(2022)]{huang2022modality}
Huang, Y., et al. (2022). Modality competition: What makes joint training of multi-modal network fail in deep learning? (Provably). \textit{Proc. ICML}, PMLR 162.

\bibitem[Ilias \& Askounis(2022)]{ilias2022multimodal}
Ilias, L. and Askounis, D. (2022). Multimodal deep learning models for detecting dementia from speech and transcripts. \textit{Front. Aging Neurosci.}, 14:830943.

\bibitem[Ivanova et al.(2022)]{ivanova2022spanish}
Ivanova, O., et al. (2022). Discriminating speech traits of Alzheimer's disease assessed through a corpus of reading task for Spanish language. \textit{Computer Speech \& Language}, 73:101341.

\bibitem[Kempler \& Goral(2010)]{kempler2010language}
Kempler, D. and Goral, M. (2010). Language and dementia: Neuropsychological aspects. \textit{Annual Rev. Applied Linguistics}, 30:87--107.

\bibitem[Kourtis et al.(2019)]{kourtis2019digital}
Kourtis, L.C., et al. (2019). Digital biomarkers for Alzheimer's disease. \textit{npj Digital Medicine}, 2:9.

\bibitem[Lanzi et al.(2023)]{lanzi2023dementiabank}
Lanzi, A.M., et al. (2023). DementiaBank: Theoretical rationale, protocol, and illustrative analyses. \textit{Am. J. Speech-Lang. Pathol.}, 32(2):426--438.

\bibitem[Lee et al.(2025)]{lee2025multimodal}
Lee, K., et al. (2025). Multimodal Alzheimer's disease recognition from image, text and audio. \textit{Scientific Reports}, 15.

\bibitem[Levelt(1989)]{levelt1989speaking}
Levelt, W.J.M. (1989). \textit{Speaking: From Intention to Articulation}. MIT Press.

\bibitem[Lin \& Washington(2024)]{lin2024scientific}
Lin, Z. and Washington, P. (2024). Multimodal deep learning for dementia classification. \textit{Scientific Reports}, 14.

\bibitem[Liu et al.(2024)]{liu2024cleverhans}
Liu, T., et al. (2024). Clever Hans effect found in automatic detection of AD through speech. \textit{Proc. Interspeech}.

\bibitem[Livingston et al.(2024)]{livingston2024lancet}
Livingston, G., et al. (2024). Dementia prevention, intervention, and care: 2024 Lancet Commission. \textit{The Lancet}, 404.

\bibitem[Luz et al.(2024)]{luz2024adressm}
Luz, S., et al. (2024). Overview of the ADReSS-M challenge on multilingual AD recognition. \textit{IEEE Open J. Signal Process.}, 5.

\bibitem[Marshall et al.(2011)]{marshall2011executive}
Marshall, G.A., et al. (2011). Executive function and instrumental activities of daily living in MCI and Alzheimer's disease. \textit{Alzheimer's \& Dementia}, 7(3):300--308.

\bibitem[Meil\'{a}n et al.(2020)]{meilan2020rhythm}
Meil\'{a}n, J.J.G., et al. (2020). Changes in speech rhythm in nondegenerative MCI and preclinical dementia. \textit{Behavioural Neurology}, 2020.

\bibitem[Mueller et al.(2018)]{mueller2018connected}
Mueller, K.D., et al. (2018). Declines in connected language are associated with very early mild cognitive impairment. \textit{Front. Aging Neurosci.}, 9:437.

\bibitem[Nichols et al.(2022)]{nichols2022gbd}
Nichols, E., et al. (2022). Estimation of global dementia prevalence in 2019 and forecast for 2050. \textit{Lancet Public Health}, 7(2).

\bibitem[Pan et al.(2023)]{pan2023path}
Pan, Y., et al. (2023). A path signature approach for speech-based dementia detection. \textit{IEEE Signal Process. Lett.}, 30.

\bibitem[Parlak(2023)]{parlak2023voice}
Parlak, D. (2023). Voice analysis in individuals with Alzheimer's disease. \textit{Brain and Behavior}, 13(10):e3271.

\bibitem[P\'{e}rez-Toro et al.(2025)]{pereztoro2025crosslingual}
P\'{e}rez-Toro, P.A., et al. (2025). Automated speech markers of Alzheimer dementia: Test of cross-linguistic generalizability. \textit{JMIR}.

\bibitem[Qi et al.(2025)]{qi2025landscape}
Qi, M., et al. (2025). AD digital biomarkers landscape and AI model scoping review. \textit{npj Digital Medicine}, 8.

\bibitem[Razani et al.(2007)]{razani2007executive}
Razani, J., et al. (2007). The relationship between executive functioning and activities of daily living in patients with relatively mild dementia. \textit{Applied Neuropsychology}, 14(4):208--214.

\bibitem[Rusko et al.(2024)]{rusko2024slovak}
Rusko, M., et al. (2024). Slovak database of speech affected by neurodegenerative diseases. \textit{Scientific Data}, 11:1320.

\bibitem[Shakeri et al.(2025)]{shakeri2025nlp}
Shakeri, H., et al. (2025). NLP in Alzheimer's disease research: Systematic review. \textit{Alz. \& Dem.: DADM}, 17(1):e70082.

\bibitem[SIDE-AD(2024)]{side2024longitudinal}
de la Fuente Garcia, S., et al. (2024). SIDE-AD: Longitudinal observational cohort study. \textit{BMJ Open}.

\bibitem[Syed et al.(2020)]{syed2020static}
Syed, Z., et al. (2020). Static vs.\ dynamic modelling of acoustic speech features for detection of dementia. \textit{IJACSA}, 11(10).

\bibitem[Ugwu \& Oyeleke(2025)]{ugwu2025taispeech}
Ugwu, C. and Oyeleke, A. (2025). Temporal-aware iterative speech model for dementia detection. \textit{arXiv:2510.00030}.

\bibitem[van den Berg et al.(2024)]{vandenberg2024remote}
van den Berg, E., et al. (2024). Digital remote assessment of speech acoustics: feasibility, reliability and associations with amyloid pathology. \textit{Alzheimer's Res. \& Therapy}, 16:1543.

\bibitem[Wang et al.(2020)]{wang2020cvpr}
Wang, W., et al. (2020). What makes training multi-modal classification networks hard? \textit{Proc. CVPR}.

\bibitem[WHO(2021)]{who2021dementia}
World Health Organization. (2021). Global status report on the public health response to dementia.

\end{thebibliography}

\end{document}